\documentstyle[graphics,epsfig,floats,aps]{revtex}

\begin{document}

\draft
\catcode`\@=11 \catcode`\@=12
\twocolumn[\hsize\textwidth\columnwidth\hsize\csname@twocolumnfalse\endcsname
\title{Gossamer supercoductivity and the mean field approximation of
a new effective Hubbard model}
\author{Yue Yu}
\address{Institute of Theoretical Physics, Chinese Academy of
Sciences, P.O. Box 2735, Beijing 100080, China}

\date{\today}

\maketitle

\begin{abstract}
We construct a new effective two-dimensional Hubbard model by
taking the different electron occupancy on site into account. The
mean field state of the new Hamiltonian gives rise to the gossamer
superconducting state proposed by Laughlin recently \cite{laugh}.
\end{abstract}

]

Since the high temperature superconductor of Cu-O cuprate was
discovered, the mechanism of the superconductivity has been
attracting much research interesting due to various unusual
properties of the high $T_c$ superconductor. A well-known model
trying to describe the normal state properties is the
2-dimensional Hubbard model or t-J model \cite{and,zhangr}. Many
analytical and numerical investigations on these two models have
been carried out \cite{ref}. Anderson proposed a resonant valence
bond (RVB) state as a mean field state of the models to explain
the undoped cuprate as a Mott insulator while the superconductor
is thought as a doped Mott insulator \cite{anderson}. Although
many progresses have been made along the clues of the Hubbard
model and t-J model, there is no persuasive evidence that the
theory of the high $T_c$ superconductivity of the cuprates can
firmly based on these two models and various extended models of
them. Especially, the quantum antiferromagnetic state may always
be favorable in these two-dimensional models for a small doping.

In a recent work, Laughlin proposed a new scenario for the high
$T_c$ superconducting cuprate, called the {\it gossamer
superconductivity} \cite{laugh}. The basic notion is that, instead
of the full projection in the RVB state, which forbits the
double-occupancy of electrons on a site, a partial projection
acting on a BCS-type wave function is introduced. If the partial
projection is not far from the full projection, one has a very
tiny superfluid density. This is called the gossamer
superconducitivity. Laughlin's explicit microscopic wave function
for the gossamer superconductor reads
\begin{eqnarray}
&&|\Psi_G\rangle=\prod_i(1-\alpha n_{i\uparrow}n_{i\downarrow})
|\Psi_{\rm BCS}\rangle,\\ &&|\Psi_{\rm BCS}\rangle=\prod_{\bf k}
(u_{\bf k}+v_{\bf k}c^\dagger_{{\bf k}\uparrow}c^\dagger_{-{\bf
k}\downarrow}|0\rangle,\nonumber
\end{eqnarray}
where $c^\dagger_{{\bf k}\sigma}$ is the Fourier component of the
electron creation operator $c^\dagger_{i\sigma}$ on a
two-dimensional lattice;
$n_{i\sigma}=c^\dagger_{i\sigma}c_{i\sigma}$ is the electron
number operator at site $i$. The projection operator
$\Pi_\alpha=\prod_i(1-\alpha n_{i\uparrow}n_{i\downarrow})$ for $
0\leq\alpha\leq 1$ is the partial Gutzwiller projection which has
an inverse $\Pi_\alpha^{-1}=\prod_i(1+\beta
n_{i\uparrow}n_{i\downarrow})$ if $\alpha<1$ and
$\beta=\alpha/(1-\alpha)$. The BCS superconducting state
$|\Psi_{\rm BCS}\rangle$ is defined as usual by
\begin{eqnarray}
&&u_k^2=\frac{1}{2}(1+\frac{\xi_{\bf k}}{E_{\bf k}}), \nonumber\\
&& v_k^2=\frac{1}{2}(1-\frac{\xi_{\bf k}}{E_{\bf k}}),\\ && u_{\bf
k}v_{\bf k}=\frac{\Delta_{\bf k}}{2E_{\bf k}},\nonumber
\end{eqnarray}
where $\xi_{\bf k}=\epsilon_{\bf k}-\mu$ and $E_{\bf k}=
\sqrt{\xi_{\bf k}^2+\Delta_{\bf k}^2}$ for the electron dispersion
$\epsilon_{\bf k}$, chemical potential $\mu$ and the
superconducting gap $\Delta_{\bf k}$. It has been pointed out that
the state $|\Psi_G\rangle$ may be superconducting even at half
filling. Furthermore, Laughlin has found that the state
$|\Psi_G\rangle$ is the exact ground state of the model
Hamiltonian
\begin{eqnarray}
H_G=\sum_{\bf k}E_{\bf k}\tilde b^\dagger_{{\bf k}\sigma}\tilde
b_{{\bf k}\sigma}, \label{goh}
\end{eqnarray}
where $\tilde b_{{\bf k}\sigma}=\Pi_\alpha b_{{\bf
k}\sigma}\Pi^{-1}$ for $ b_{{\bf k}\uparrow}=u_{\bf k} c_{{\bf
k}\uparrow}+v_{\bf k} c_{-{\bf k}\downarrow}$ and $ b_{{\bf
k}\downarrow}=u_{\bf k} c_{{\bf k}\downarrow}-v_{\bf k} c_{-{\bf
k}\uparrow}$ annihilate the BCS state. Explicitly, $\tilde b_{{\bf
k}\sigma}$ reads
\begin{eqnarray}
\tilde b_{{\bf k}\uparrow}&=&\frac{1}{\sqrt N}\sum_je^{i{\bf
k}\cdot{\bf r}_j}[u_{\bf k}(1+\beta
n_{j\downarrow})c_{j\uparrow}+v_{\bf k}(1-\alpha
n_{j\uparrow})c^\dagger_{j\downarrow}],\nonumber\\ \tilde b_{{\bf
k}\downarrow}&=&\frac{1}{\sqrt N}\sum_je^{i{\bf k}\cdot{\bf
r}_j}[u_{\bf k}(1+\beta n_{j\uparrow})c_{j\downarrow}-v_{\bf
k}(1-\alpha n_{j\downarrow})c^\dagger_{j\uparrow}].\nonumber\\
\label{deb}
\end{eqnarray}
Laughlin has shown that for any magnitude of $\alpha$, the
quasiparticle energies remain at $E_{\bf k}$, which indicates the
psuedogap phenomenon. Right following up Laughlin's work, Zhang
has checked the gossamer superconductivity in a more realistic
effective Hubbard model \cite{zhang}. It was found that the
gossamer superconducting state is similar to the RVB
superconducting state, except that the chemical potential is
approximately pinned at the mid of the two Hubbard bands away from
the half filling.

Ref.\cite{zhang} indicated that the gossamer superconducting state
may possibly be a good variational state of the effective Hubbard
model. Nevertheless, it is still in question to make an explicit
relation between the Hamiltonian (\ref{goh}) and that of the
effective Hubbard model used in ref. \cite{zhang}. In this paper,
we try to provide a new effective Hubbard model which is the
extension of the Hubbard model and t-J model. Our extending method
is other than all the previous extended models of those two
models. We consider the case that the lattice sites are allowed to
be double-occupied by electrons with opposite spins. Because of
the on-site Coulomb repulsion between electrons, the electrons
hopping between sites becomes dependent on the occupancy of sites.
There are three possible hopping processes that may have different
hopping probabilities (see, Fig. 1). On the other hand, when
electrons are paired, the coupling between pairs may also be
dependent on the the occupancy of the sites (Fig. 2) while the
hopping may be dependent on the paring gap. What is new by
employing such a new effective Hubbard model is that the Laughlin
gossamer superconducting state may be viewed as the mean field
state of this model. Namely, there is a gossamer superconducting
phase in the system described by this new effective Hubbard
Hamiltonian.

To extend the Hubbard model, we first write down the Hamiltonian
of the effective Hubbard model
\begin{eqnarray}
H_{eh}&=&-\sum_{i\ne
j;\sigma}(t_{ij}c^\dagger_{i\sigma}c_{j\sigma} +h.c)+U\sum_i
n_{i\uparrow}n_{i\downarrow}\nonumber\\&-&\mu\sum_{i,\sigma}
n_{i\sigma}+ \sum_{i\ne j} J_{ij}({\bf S}_i\cdot {\bf
S}_j-\frac{1}4n_in_j) , \label{hh}
\end{eqnarray}
where we have included the chemical potential term and ${\bf
S}=\frac{1}{2}c^\dagger_\sigma{\bf
\tau}_{\sigma\sigma'}c_{\sigma'}$ is the local spin operator. One
can rewrite this Hamiltonian as
\begin{eqnarray}
H_{eh}&=&-\sum_{i\ne
j;\sigma}(t_{ij}c^\dagger_{i\sigma}c_{j\sigma}
+h.c)-\frac{1}2\sum_{i\ne j}
\hat{D}^\dagger_{ij}J_{ij}\hat{D}_{ij}\nonumber\\&+& U\sum_i
n_{i\uparrow}n_{i\downarrow}-\mu\sum_{i,\sigma} n_{i\sigma},
\label{hhr}
\end{eqnarray}
where $\hat{D}_{ij}=c_{i\downarrow}c_{j\uparrow}-c_{i\uparrow}
c_{j\downarrow}$ is the pairing operator. Now, we extend this
model according to our consideration mentioned above. Due to the
different occupant situations as shown in Fig. 1, one can extend
$t_{ij}$ to be an operator $\hat t_{ij,\sigma}$:
\begin{eqnarray}
\hat t_{ij,\sigma}=t_{ij}+ t^{(1)}_{ij} n_{i\bar\sigma}+
t^{(2)}_{ij}n_{i\bar\sigma}n_{j\bar\sigma},
\end{eqnarray}
where the first term is the common site-dependent hopping
probability; the second term is corresponding to the correction to
the hopping process for the occupancy like Fig. 1(b) and the third
term to that like Fig. 1(c). This kind of 'hopping' terms has been
met when we deal with the magnetic impurity problem \cite{yu}. The
exact version of $ t^{(a)}_{ij}$ via a microscopic calculation is
not obtained in the present work. However, we can estimate them
through the following physical consideration. It is known that
$t_{ij}=-\frac{1}{N}\sum_{\bf k}\epsilon_{\bf k}e^{i{\bf
k}\cdot({\bf r}_i-{\bf r}_j)}$. Notice that the chemical potential
and the gap-dependence, we can estimate $\delta^{(a)} t_{ij}$ by
\begin{eqnarray}
&& t^{(1)}_{ij}=\frac{\alpha}{N}\sum_{\bf k}[a_1\xi_{\bf
k}+b_1(E_{\bf k}-\xi_{\bf k})]e^{i{\bf k}\cdot({\bf r}_i-{\bf
r}_j)},\label{corr}\\ && t^{(2)}_{ij}=\frac{\alpha^2}{N}\sum_{\bf
k}[a_2\xi_{\bf k}+b_2(E_{\bf k}-\xi_{\bf k})]e^{i{\bf k}\cdot({\bf
r}_i-{\bf r}_j)},\nonumber
\end{eqnarray}
where the coefficients $a_l$ and $b_l$, in principle, depend on
$U$ and $J_{ij}$ but we are unable to calculate them at present.
We leave them to be determined later. The $\alpha$-dependence in
eq. (\ref{corr}) is because in the limit $U\to 0$, $\alpha\to 0$
while the hopping probability becomes independent of the
site-occupancy.

The coupling between the pairing operators $\hat{D}_{ij}$ and
$\hat{D}^\dagger_{ij}$ may also dependent on the occupancy of the
sites $i$ and $j$ (See Fig. 2):
\begin{eqnarray}
\hat J_{ij}=J_{ij}+J^{(1)}_{ij}(n_i+n_j)+J^{(2)}_{ij}n_in_j,
\end{eqnarray}
where
\begin{eqnarray}
 J^{(1)}_{ij}=A_1\alpha
J_{ij},~~ J^{(2)}_{ij}=A_2\alpha^2 J_{ij},
\end{eqnarray}
with the undetermined coefficients $A_l$. The extended effective
Hubbard model we are considering is given by
\begin{eqnarray}
H_{\rm eff}&=&-\sum_{i\ne
j;\sigma}(\hat{t}_{ij,\sigma}c^\dagger_{i\sigma}c_{j\sigma}
+h.c)-\frac{1}2\sum_{i\ne j}
\hat{D}^\dagger_{ij}\hat{J}_{ij}\hat{D}_{ij}\nonumber\\&+&
 +U\sum_i
n_{i\uparrow}n_{i\downarrow}-\mu\sum_{i,\sigma} n_{i\sigma}.
\label{heff}
\end{eqnarray}

To see the gossamer superconducting state, we consider the mean
field state of the Hamiltonian (\ref{heff}). Using the gap order
parameter $D_{ij}=\langle \hat{D}_{ij}\rangle$ to replace the
pairing operator $\hat{D}_{ij}$ in the Hamiltonian (\ref{heff}),
one has, up to a chemical potential re-definition, the mean field
Hamiltonian is given by
\begin{eqnarray}
H_{\rm MF}&=&-\sum_{i\ne
j;\sigma}(\hat{t}_{ij,\sigma}c^\dagger_{i\sigma}c_{j\sigma}
+h.c)-\frac{1}2\sum_{i\ne j}
\hat{\Delta}^\dagger_{ij}\hat{D}_{ij}\nonumber\\
&-&\frac{1}2\sum_{i\ne j} \hat{D}^\dagger_{ij}\hat{\Delta}_{ij}
+U\sum_i n_{i\uparrow}n_{i\downarrow}-\mu_R\sum_{i,\sigma}
n_{i\sigma}, \label{MF}
\end{eqnarray}
where $\hat\Delta\dagger_{ij}=D^*_{ij}\hat{J}_{ij}$. Now we make a
special choice of the parameters such that
\begin{eqnarray}
&& a_1=2\beta/\alpha,~~~ b_1=(\alpha+\beta)/\alpha,\nonumber\\ &&
 a_2=\frac{\alpha^2-\beta^2}{2\alpha^2},~~~b_2=-\frac{\alpha^2+\beta^2}
 {2\alpha^2},\label{p1}\\
 && J^{(1)}_{ij}=(\alpha+\beta)J_{ij},~~~J^{(2)}_{ij}=
 \alpha\beta J_{ij},\nonumber
 \end{eqnarray}
 and $U=U_G$ and $\mu_R=\mu+\mu_G$ with
 \begin{eqnarray}
 && U_G=\frac{1}{2N}\sum_{\bf k}[(\beta-\alpha)E_{\bf
 k}+(\beta+\alpha)\xi_{\bf k}],\nonumber\\
 && \mu_G=\frac{1}{N}\sum_{\bf k}[2(\alpha-1)\xi_{\bf k}+2\alpha
 E_{\bf k}].\label{p2}
 \end{eqnarray}
 Substituting eqs. (\ref{p1}) and (\ref{p2}) into $H_{\rm MF}$,
 one has
 \begin{eqnarray}
 H^G_{{\rm MF}}&=&-\sum_{i\ne j;\sigma}(\hat{t}_{ij,\sigma}^G c^\dagger_{i\sigma}
 c_{j\sigma}+h.c.)+U_G\sum_in_{i\uparrow}n_{i\downarrow}\nonumber\\
 &+&\sum_{i\ne
 j;\sigma}(-1)^\sigma(\hat{\Delta}_{ij,\sigma}c^\dagger_{i\sigma}
 c^\dagger_{j\bar\sigma}+h.c.)-\mu_G\sum_{i\sigma}n_{i\sigma},\label{MFG}
\end{eqnarray}
where
\begin{eqnarray}
\hat{t}^G_{ij,\sigma}&=&\sum_{\bf k}\frac{E_{\bf
k}}{N}[\frac{1}{2}(v_{\bf k}^2-u_{\bf k}^2)-(\alpha v_{\bf
k}^2+\beta u_{\bf k}^2)n_{i\bar\sigma}\nonumber\\
&+&\frac{1}{2}(\alpha^2v_{\bf k}^2-\beta^2u_{\bf
k}^2)n_{i\bar\sigma}n_{j\bar\sigma}]e^{i{\bf k}\cdot({\bf
r}_i-{\bf r}_j)}, \\ \hat{\Delta}_{ij,\sigma}&=&\sum_{\bf
k}\frac{E_{\bf k}}Nu_{\bf k}v_{\bf k}e^{i{\bf k}\cdot({\bf
r}_i-{\bf r}_j)}\nonumber\\ &\times&
(1+(\alpha+\beta)n_{i\bar\sigma}+\alpha\beta
n_{i\bar\sigma}n_{j\sigma}).\nonumber
\end{eqnarray}
Substituting the definition (\ref{deb}) of $\tilde b_{{\bf
k}\sigma}$ into the gossamer superconducting Hamiltonian
(\ref{goh}), one can directly check that the gossamer
superconducting Hamiltonian (\ref{goh}) is exactly the same as the
mean field Hamiltonian (\ref{MFG}), i.e.,
\begin{eqnarray}
H_{\rm MF}^G=H_G.
\end{eqnarray}
Thus, we find that the gossamer superconducting state is indeed a
fixed point of the system described by the effective Hubbard model
(\ref{heff}). If the parameters of the hopping, exchange, chemical
potential and interaction are not far from their fixed point
values, the system may exhibit a gossamer superconductity.

In conclusions, we have constructed a new effective Hubbard model
in which the different hopping probabilities and the pairing
couplings are introduced due to the different site-occupancy at
sites. We showed that there is a superconducting phase in such a
system since the mean field Hamiltonian of this system in a proper
choice of the parameters is just the gossamer superconducting
Hamiltonian. Beside the superconducting phase, this new effective
Hubbard model is anticipated to have a fruitful phase structure.
In ref. \cite{zhang}, it was shown that there is a critical
interaction strong $U_c$ that separates the gossamer
superconducting phase from the Mott insultor phase. In this new
effective Hubbard model, we see that even the superconducting gap
vanishes or the superfluid density is completely suppressed
($\alpha=1$), the dispersion of electron as well as the
interaction between electrons for the latter are dependent on the
occupancy of the site that electrons lying on. We expect these
unusual behaviors of the electrons may cause unusual normal state
properties and relate to the anomalous features in the cuprates.
It is possible that the phase diagram of this model may have a
better overlap to the cuprate superconductors. The further works
are in progress.

The author is grateful to useful discussions with Shaojin Qin, Tao
Xiang and Lu Yu. He would like to thank Shiping Feng for him to
draw the author's attention to gossamer superconductivity. This
work was partially supported by the NSF of China.

\noindent{Fig.1 Three possible hopping processes. (a) An electron
at a single-occupant site hops to an empty site; (b) An electron
at a double-occupant site hops to an empty site; (c)  An electron
with spin $\sigma$ at a double-occupant site hops to a site which
is occupied by a spin $\bar\sigma$.}

\vspace{2mm}

\noindent{Fig.2 Three possible electron pairs depending on the
site occupancy. (a) the pairs with both sites single-occupied; (b)
the pairs with one site single-occupied and another
double-occupied; (c) the pairs with both sites double-occupied.}

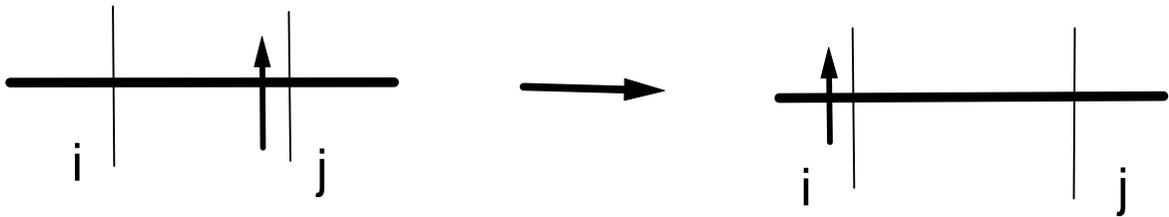
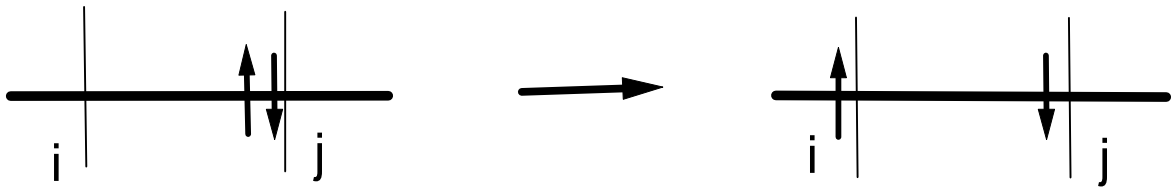
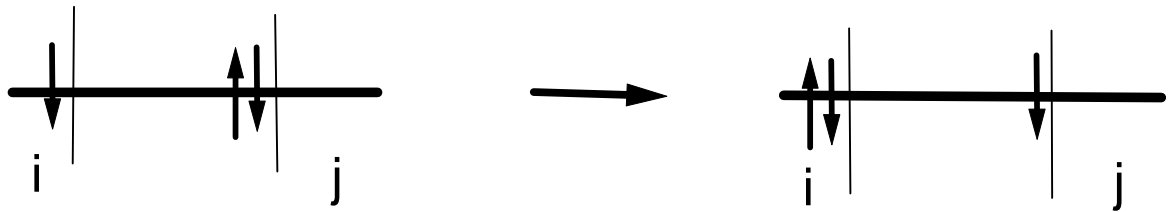
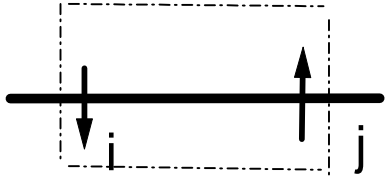
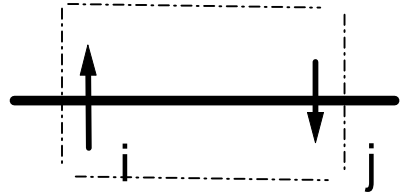
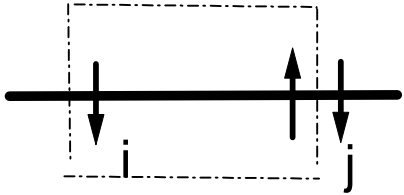
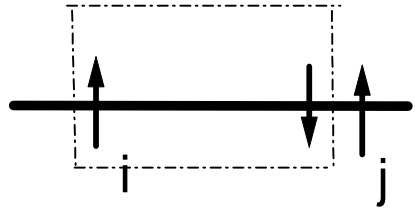
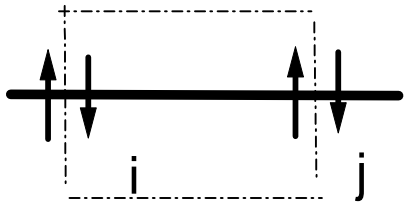
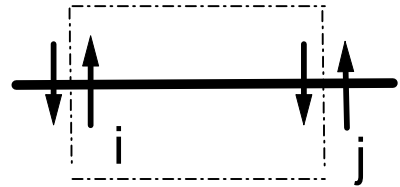
\end{document}